\documentclass[prl,amsmath,amssymb,twocolumn,superscriptaddress]{revtex4-2}
\usepackage{amsmath}
\usepackage{amssymb}
\usepackage{amstext}
\usepackage{amsfonts}
\usepackage{amsxtra}
\usepackage{bm}
\usepackage[usenames]{color}
\usepackage{grffile}
\usepackage{soul}
\usepackage{epstopdf}
\usepackage{xcolor}
\usepackage{graphicx}
\usepackage{marvosym}
\usepackage{wasysym}
\usepackage{hyperref}
\usepackage{orcidlink}

\hypersetup{colorlinks=true, linkcolor=black, citecolor=blue, urlcolor=blue}

{%
\setlength{\fboxsep}{0pt}%
\setlength{\fboxrule}{1pt}%
}%
\begin{document}


\title{Quantum vortex channels as Josephson junctions}

\author{Natalia Masalaeva\,\orcidlink{0000-0002-9314-8973}}
\email[Corresponding author: ]{natalia.masalaeva@uibk.ac.at}
 \affiliation{
     Universit\"{a}t Innsbruck, Fakult\"{a}t f\"{u}r Mathematik, Informatik und Physik, Institut f\"{u}r Experimentalphysik, 6020 Innsbruck, Austria}
 
\author{Wyatt Kirkby}
 \affiliation{
 	Kirchhoff-Institut f\"{u}r Physik, Universit\"{a}t Heidelberg, Im Neuenheimer Feld 227, 69120 Heidelberg, Germany}
 	\affiliation{Physikalisches Institut, Universit\"{a}t Heidelberg, Im Neuenheimer Feld 226, 69120 Heidelberg, Germany}
 
\author{Francesca Ferlaino\,\orcidlink{0000-0002-3020-6291}}
 \affiliation{
 	Universit\"{a}t Innsbruck, Fakult\"{a}t f\"{u}r Mathematik, Informatik und Physik, Institut f\"{u}r Experimentalphysik, 6020 Innsbruck, Austria}
 	\affiliation{
 	Institut f\"{u}r Quantenoptik und Quanteninformation, \"{O}sterreichische Akademie der Wissenschaften, Innsbruck 6020, Austria}
 
\author{Russell N. Bisset}
\email[Corresponding author: ]{russell.bisset@uibk.ac.at}
 \affiliation{
 	Universit\"{a}t Innsbruck, Fakult\"{a}t f\"{u}r Mathematik, Informatik und Physik, Institut f\"{u}r Experimentalphysik, 6020 Innsbruck, Austria}

\begin{abstract}
 
In quantum gases, weak links are typically realized with externally imposed optical potentials.
We show that, in rotating binary condensates, quantized vortices in one component form hollow channels that act as self-induced weak links for the other, enabling superflow through otherwise impenetrable, phase-separated domains. 
This introduces a novel barrier mechanism: quantum pressure creates an effective barrier inside the vortex channel, set by the constriction width, which controls the superflow.
Tuning the interspecies interaction strength drives a crossover from the hydrodynamic transport to Josephson tunneling regime. 
Long-range dipolar interactions further tune the weak-link properties, enabling both short links and two coupled junctions in series.
Circuit models quantitatively capture the dc current-phase relations for both configurations.
These results establish vortices as reconfigurable, interaction-controlled Josephson elements in superfluids.

\end{abstract}

\date{\today}
\maketitle

\emph{Introduction}---Quantum transport through a weak link---such as a localized constriction or barrier between two superconducting or superfluid reservoirs \cite{Likharev1979,Zibold2019}---is fundamentally characterized by the current-phase relation (CPR), which describes the dependence of the supercurrent on the phase difference across the link \cite{Golubov2004}. Traditionally, weak links are realized in superconductors using thin insulating barriers \cite{Anderson1963,Shapiro1963} or Dayem microbridges \cite{Dayem1967}, in superfluid helium using arrays of microapertures \cite{Backhaus1997,Hoskinson2005}, and in ultracold gases using optical barriers \cite{Cataliotti2001,Albiez2005,Levy2007,LeBlanc2011,Brantut2012}.
These platforms have revealed the hallmarks of the Josephson effect, including sinusoidal and multivalued CPRs, hysteresis loops, and quantized phase slips \cite{Barone1982,Davis2002superfluid,Gati2007,Amico2021,Polo2025}.

In ultracold gases, optical potentials have been widely used to engineer weak links of various geometries \cite{Brantut2013,Krinner2014,Husmann2015,Spagnolli2017,Fabritius2024}, enabling studies of Josephson physics, including dc and ac Josephson effects \cite{Levy2007,Valtolina2015,Kwon2020}, CPRs \cite{Luick2020,Singh2025} and Shapiro steps \cite{DelPace2025,Bernhart2025}.
In addition, toroidal quantum gases with localized repulsive barriers have allowed the creation of long-lived persistent currents \cite{Ramanathan2011,Wright2022,Pace2022}, investigations of their stability and phase slips \cite{Moulder2012,Wright2013driving,Murray2013,Minguzzi2019,Pezz2024}, and measurements of the CPR \cite{Eckel2014}.

More recently, attention has turned to self-induced weak links, which arise from internal interactions rather than external potentials.
In dipolar supersolids, self-organized density modulations form barriers that act as Josephson junctions \cite{Ilzhfer2021,Biagioni2024,Pezze2025}, enabling coherent tunneling without applied structuring.

\begin{figure}
	\centering	\includegraphics[width=0.48\textwidth]{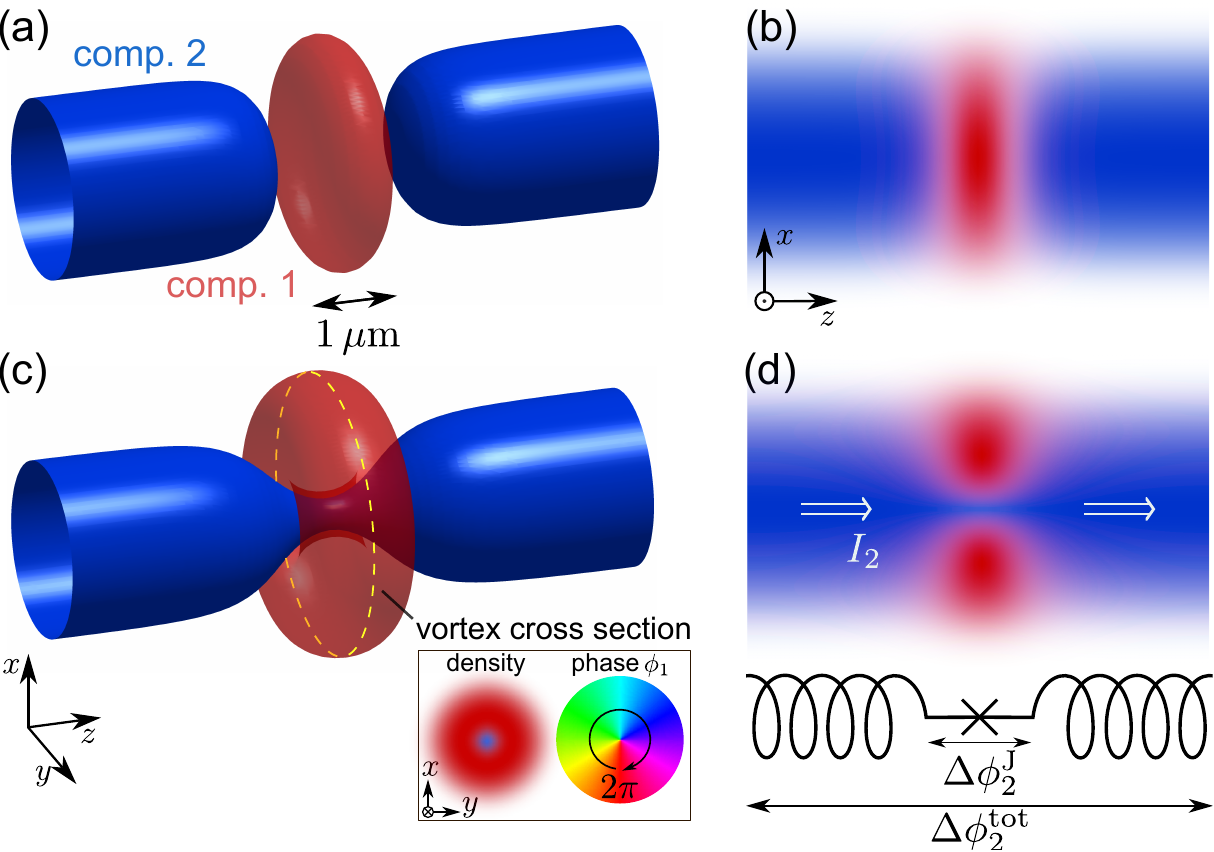}
	\caption{
	Vortex-mediated weak link. (a,b) Vortex-free immiscible state: component 1 (red, dipolar) blocks current in component 2 (blue, nondipolar).
	(c,d) A vortex oriented along $z$ resides in component 1; component 2 flows through its hollow core, carrying current $I_2$.
	Magnetostriction produces thin dipolar domains, yielding a short weak link \cite{Kirkby2023Spin}.
	Density isosurfaces of comp.~1 (comp.~2) are shown at 45\% (70\%) of their respective peak densities.
	 (d, lower): circuit model: an inductor in series with a nonlinear Josephson junction (cross), with total phase drop $\Delta\phi^{\mathrm{tot}}_{2}$ and junction phase drop $\Delta\phi^{\rm{J}}_{2}$. Parameters: interspecies scattering length $a_{12}=125\,a_0$; populations $N_2=4N_1=7.5\times 10^3$.
	}
	\label{fig:Model}
\end{figure}

In this Letter, we demonstrate that rotation-induced vortices create self-organized weak links in a binary Bose-Einstein condensate (BEC), eliminating the need for externally imposed optical potentials.
In the absence of rotation [Fig.~\ref{fig:Model}(a,b)], repulsive intercomponent interactions drive phase separation: the red component forms an impenetrable domain that blocks the flow of the blue component.
With a quantized vortex in the red component [Fig.~\ref{fig:Model}(c,d)], the vortex core is hollow---free of the red component---and becomes a channel through which the blue component flows, carrying current $I_2$.
This channel realizes a new barrier mechanism---a quantum-pressure barrier set by radial confinement in the vortex core. Its height and effective length are tunable by intra- and intercomponent interactions (including dipole-dipole coupling), and component populations, enabling precise interaction-based control in a self-organized geometry.

From a practical perspective, vortices are readily generated by established methods---trap rotation \cite{madison2000vortex,abo2001observation,haljan2001driving}, phase imprinting in binary condensates \cite{Matthews1999,Anderson2000,williams1999preparing}, and magnetostirring in dipolar gases \cite{klaus2022observation}.
Under steady rotation, vortex lattices form; each core then acts as a weak link.
While we analyze a single core to expose the elementary CPR, the mechanism extends directly to lattices and multi-vortex configurations, providing weak-link arrays analogous to aperture arrays in superfluid helium \cite{Backhaus1997,Hoskinson2005}.

By analyzing the CPR of the vortex-core channel, we show that rotation-induced vortices function as Josephson weak links.
As the interspecies scattering length increases, the CPR crosses over from nearly linear (hydrodynamic) to sinusoidal (tunneling) \cite{Hoskinson2005,LeBlanc2011}.
A minimal circuit model---an ideal Josephson element in series with a kinetic inductance [see bottom panel of Fig.~\ref{fig:Model}(d)]---captures both regimes; importantly, the inductance is calculated independently and is not treated as a fit parameter.
When the vortex channel length increases, nonlocal interactions reshape the core into two coupled junctions, for which an extended circuit model recovers quantitative agreement.
These results establish vortex-mediated links as a versatile platform for exploring Josephson dynamics in atomtronic circuits.

\emph{System}---Each component $i=1,2$ of the binary dipolar BEC is described by order parameter $\psi_i=\sqrt{n_i}\,e^{i\phi_i}$ with density $n_i$ and phase $\phi_i$.
Such mixtures are now routinely realized in experiments~\cite{Trautmann2018,Ravensbergen2020,Schafer2023,lecomte2025production,xie2025feshbach,kalia2025creation}.
The system is confined in a three-dimensional (3D) tube, with harmonic transverse confinement $\omega_x=\omega_y=2\pi\times 100\ \mathrm{s}^{-1}$ and an untrapped axial direction.
We consider a dipolar-nondipolar mixture ($^{162}\mathrm{Dy}$ + nondipolar), with magnetic moments $\mu_1=9.93\,\mu_B$ and $\mu_2 = 0$.
Contact interactions are $a_{11}=a_{22}=130\,a_0$, while the interspecies scattering length $a_{12}$ is used as a control parameter.

Our system is in the immiscible regime [Fig.~\ref{fig:Model}(a,b)], achieved by tuning $a_{12}$.
To illustrate vortex-mediated weak links, we focus on two configurations: a vortex in the dipolar component [Fig.~\ref{fig:Model}(c,d)] produces a short weak link because magnetostriction axially compresses the dipolar domain hosting the vortex \cite{Kirkby2023Spin}, whereas a vortex in the non-dipolar component generates a long, narrow channel.
Although the ground states exhibit no superfluid flow, a dc current can be induced by imprinting a phase gradient on the component that fills the vortex core.
We evolve the coupled dipolar Gross-Pitaevskii equations in imaginary time at fixed phase bias, allowing the system to relax into a stationary dc state (see End Matter for details). 
The vortex core thus forms a hollow channel that permits one superfluid to penetrate through the domains of the other [Fig.~\ref{fig:Model}(c,d)].
These constrictions resemble Dayem microbridges in superconductors \cite{Dayem1967,Granata2016} and narrow apertures in liquid helium \cite{Backhaus1997,Sukhatme2001,Hoskinson2005}, but here they emerge naturally from interaction effects \cite{Abad2011dipolar,Abad2011,Tengstrand2021,Tengstrand2023,Bellettini2024,Jezek2025}, rather than from external structuring.

\emph{Current-phase relation}---The CPR describes the dependence of the current on the phase difference across the system, $\Delta\phi^{\rm{tot}}_i = \phi_i(\rho = 0,l/2)-\phi_i(\rho = 0,-l/2)$, where $\rho = \sqrt{x^2+y^2}$ and $l$ is the system length; radial variations of the phase are negligible owing to the system's elongated geometry.
The atomic current density is
\begin{equation}
	\mathbf{I}_i (\mathbf{x)} = n_i(\mathbf{x)} \frac{\hbar}{m_i} \nabla\phi_i\,(\mathbf{x)},
\end{equation}
where $m_i$ is the atomic mass.
In what follows, we consider the radially integrated current, $I_i = \int I_{z,i}\,(\mathbf{x})\,d^2\mathbf{\rho}$, which is independent of $z$.

Because the current-carrying component is radially constricted within the vortex channel [Figs.~\ref{fig:Model}(c,d)], it experiences a quantum-pressure-induced barrier. 
To quantify this barrier and its influence on the current, we define for component $i$ an effective potential from the radial quantum-pressure term,
\begin{equation}
  V_i^{\mathrm{eff}}(\mathbf{x}) = -\frac{\hbar^2}{2m_i}\,\frac{\nabla_\rho^2 \sqrt{n_i(\mathbf{x})}}{\sqrt{n_i(\mathbf{x})}},
  \label{eq:eff_pot}
\end{equation}
where $\nabla_\rho^2$ is the radial part of the Laplacian in cylindrical coordinates.
The barrier height is $V_{b,i}=  \max\big[V_i^{\mathrm{eff}}(\rho=0,z)\big]$. 
Unlike in quantum-transport experiments, where the potential landscape is externally engineered \cite{Krinner2017}, here $V_i^{\mathrm{eff}}$ is set by the controllable vortex-channel width $w$ \newcounter{widthfn}\footnote{We define $w$ as the minimum, over $z$, of the half-width at half-maximum of the radial density profile of the component that fills the vortex core.}\setcounter{widthfn}{\value{footnote}}, which in turn is determined by the spin healing length $\xi_s$. The latter is highly sensitive to the interspecies scattering length and diverges near the miscibility threshold as
$\xi_s \propto |a_{12}-a_c|^{-1/2}$, with $a_c$ the miscible-immiscible transition point~\cite{Tsubota2011}.
Consequently, a Feshbach resonance on $a_{12}$ provides a natural, highly responsive knob for engineering weak-link properties.

\begin{figure}[t!]
	\centering
	\includegraphics[width=0.48\textwidth]{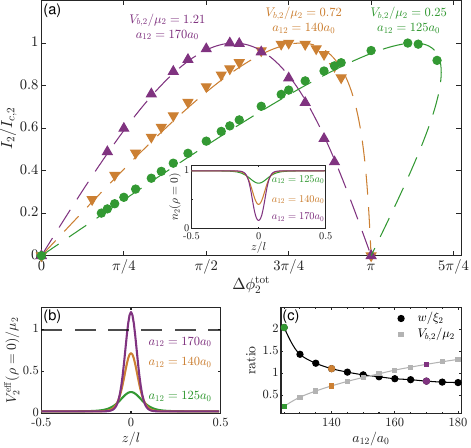}
	\caption{
		Short vortex junction. (a) CPRs for different $a_{12}$ (barrier heights). Markers represent numerical data and dashed lines are the circuit model.
		Each CPR is normalized to its critical (peak) current $I_{c,2}$.
		Inset shows the corresponding core densities normalized to their peaks.
		(b) Effective potential inside the vortex core.
		(c) Vortex channel width $w$ (black) and barrier height $V_{b,2}$ (gray) versus interspecies scattering length. 
		}
	\label{fig:Short}
\end{figure}

\emph{Short vortex junction}---We first consider the junction configuration in which a vortex is imprinted in component 1 (dipolar) [Fig.~\ref{fig:Model}(c,d)].
The bulk healing length of component 2 (current carrying), $\xi_2=0.35\,\mu m$~\footnote{We define $\xi_2 = \hbar/\sqrt{m_2 g_{22}n_{2,\rm{max}}}$, where $g_{22}$ is an intraspecies contact interaction strength (see End Matter), and $n_{2,\rm{max}}$ is the peak 3D density}, is of the same order of magnitude as the vortex length along $z$, $l_v\approx 1.1\,\mu m$ \footnote{See End Matter (last section) for the criteria used to determine the vortex length and the junction region.}, which we refer to this as the ``short junction'' regime~\cite{Margineda2023}.
The CPRs at three values of the interspecies scattering length $a_{12}$ are shown in Fig.~\ref{fig:Short}(a), with the corresponding effective barriers in Fig.~\ref{fig:Short}(b).
As $a_{12}$ increases, the CPR crosses over from nearly linear (hydrodynamic; green dots) to sinusoidal (purple triangles), indicating the onset of Josephson tunneling through a weak link.

In the hydrodynamic regime ($a_{12}=125\,a_0$), the vortex core is broad ($w/\xi_2 \approx 2$), reducing the radial curvature of the density and yielding a potential barrier well below the chemical potential $\mu_2$ [Fig.~\ref{fig:Short}(b)].
Consequently, the density along the $z$ axis exhibits only a shallow dip [Fig.~\ref{fig:Short}(a), inset].
The resulting CPR [green dots in Fig.~\ref{fig:Short}(a)] is nearly linear, $I_2 \propto \Delta\phi_2^{\mathrm{tot}}$, consistent with hydrodynamic superflow strongly coupled through the vortex channel.
Positive currents persist for $\Delta\phi_2^{\rm tot}>\pi$,  analogous to observations in superfluid $^4$He flowing through narrow apertures \cite{Hoskinson2005}; there, unlike in our system, the aperture widths were fixed and the healing length was tuned by temperature.

As $a_{12}$ increases, the system is driven deeper into the immiscible regime, narrowing the vortex channel.
Because the effective barrier scales as $w^{-2}$ [Eq.~(\ref{eq:eff_pot})], its height rises rapidly with increasing $a_{12}$.
At $a_{12}=170\,a_0$, the crossover to the Josephson tunneling regime is complete: the barrier height exceeds the chemical potential $\mu_2$ [Fig.~\ref{fig:Short}(b)], the 3D core density is strongly depleted [Fig.~\ref{fig:Short}(a), inset], and the CPR is well-described by a sinusoidal form, $I_2 \propto I_{c,2}\sin\Delta\phi_2^{\mathrm{tot}}$ [purple triangles in Fig.~\ref{fig:Short}(a)].

In Fig.~\ref{fig:Short}(c), we quantify how the vortex width and barrier height evolve with increasing $a_{12}$.
Crucially, the reduction of the width to values comparable to the healing length ($w/\xi_2 \lesssim 1$) coincides with the barrier height entering the tunneling regime ($V_{b,2}/\mu_2 \gtrsim 1$).
This crossover highlights the tunability of the weak link in our system: by adjusting $a_{12}$, both the channel width and the emergent barrier height can be controlled, enabling switching between strongly coupled (hydrodynamic) and weakly coupled (tunneling) core-current regimes.

\begin{figure*}[t!]
	\centering	
	\includegraphics[width=0.98\textwidth]{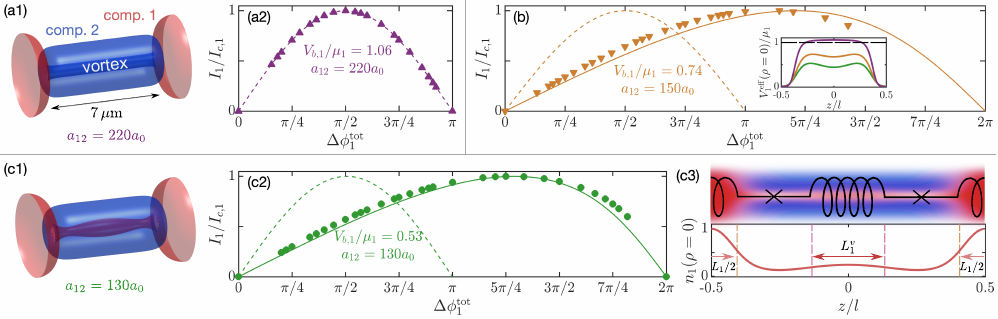}
	\caption{Long-vortex junction for three barrier strengths: (a) $a_{12}=220\,a_0$ (tunneling regime), (b) $a_{12}=150\,a_0$, and (c) $a_{12}=130\,a_0$ (hydrodynamic regime).
(a2), (b), and (c2): CPRs with numerical data (markers) shown together with the simple circuit model [Eqs.~\eqref{Eq:ModelPhase}-\eqref{L_kin}] (dashed lines) and the extended circuit model [Eq.~\eqref{Eq:extendedmodel}] (solid lines).
(b, inset) depicts the effective barrier for all three values of $a_{12}$.
(a1) shows a channel in the tunneling regime, whereas (c1) shows that in the hydrodynamic regime the dipolar interactions restructure the vortex core environment into a double junction: the flowing component (red) exhibits a density minimum near each end of the vortex separated by a density bulge in the middle. The first component is shown at 10\% of its peak density and the second at 70\%.
(c3) Extended circuit model (top) and axial density (bottom).
The local density maximum at the center of the vortex core is modeled by an additional kinetic inductance $L_1^{v}$, while the two minima are represented as Josephson elements.
	}
	\label{fig:Long}
\end{figure*}

\emph{Circuit model}---The obtained CPRs are captured by a Deaver-Pierce circuit model \cite{Deaver1972,Piazza2010,Wright2013driving}, consisting of a kinetic (hydrodynamic) inductance $L_i$ of the flowing component $i$ in series with an ideal Josephson element with a sinusoidal CPR, as illustrated at the bottom of Fig.~\ref{fig:Model}(d).
The total phase drop is the sum of the phase across the ideal weak link and across the inductive element,
\begin{equation}
\Delta\phi^{\rm{tot}}_i = \Delta\phi^{\rm J}_i + \frac{L_i I_i}{\hbar}, \label{Eq:ModelPhase}
\end{equation}
while the ideal junction obeys the Josephson relation
\begin{equation}
I_i = I_{c,i}\sin\Delta\phi^{\rm J}_i, \label{Eq:ModelCPR}
\end{equation}
where $I_{c,i}$ is extracted as the peak of the numerically determined CPR.
The kinetic inductance arises from density distribution and is given by
\begin{equation}
L_i = m_i \int_{\mathcal{R}}\!\frac{dz}{n_{i,\rm{1D}}(z)} \, ,
\label{L_kin}
\end{equation}
with the one-dimensional density $n_{i,\rm{1D}} = \int\! dx\,dy\, n_i(\mathbf{x})$ integrated over the transverse directions.
Here $\mathcal{R}$ denotes the region outside the vortex core, identified by the threshold condition $n_i (\rho = 0,z) \geqslant n_i^{\rm{min}} + f(n_i^{\rm{max}} -n_i^{\rm{min}})$, 
where $n_i^{\rm{min/max}}$ are the global minimum/maximum densities along the core ($\rho=0$), and $f = 0.5$ [see Fig.~\ref{fig:EM_induc}(a) in End Matter for visualization].

The CPRs predicted by the circuit model [Eqs.~\eqref{Eq:ModelPhase} and~\eqref{Eq:ModelCPR}] are shown as dashed lines in Fig.~\ref{fig:Short}(a).
Beyond the excellent quantitative agreement with the simulations, the model furnishes deeper intuition.
In the high-barrier tunneling regime ($a_{12}=170\,a_0$), the supercurrent is small and the phase drop across the inductive element is negligible [see Eq.~\eqref{Eq:ModelPhase}], so $\Delta\phi_i^{\mathrm{tot}} \approx \Delta\phi_i^{\rm J}$, yielding the sinusoidal CPR observed in Fig.~\ref{fig:Short}(a).
In contrast, in the low-barrier hydrodynamic regime ($a_{12}=125\,a_0$) the larger current creates a significant phase drop across the inductive element; Eq.~(\ref{Eq:ModelPhase}) thus accounts for total phase drops $\Delta\phi_2^{\mathrm{tot}} > \pi$.
The circuit model also predicts a multivalued hydrodynamic CPR, although the lower branch is absent in our numerics because these states are dynamically unstable \cite{Piazza2010}.
The excellent agreement between numerics and the circuit model supports the suitability of the Deaver-Pierce framework for capturing the interplay of kinetic inductance and Josephson nonlinearity in our system.

\emph{Long-vortex junction}---We now consider the long-junction configuration, in which the vortex resides in the second (nondipolar) component, shown in Fig.~\ref{fig:Long}(a1,c1) for two representative interspecies scattering lengths, $a_{12}=220\,a_0$ and $a_{12}=130\,a_0$.
Because domains of the nondipolar component do not undergo magnetostriction, they can be arbitrarily long \cite{Kirkby2023Spin}; placing the vortex in this component therefore realizes an elongated vortex core.
Note that single long-lived vortex lines aligned with the trap axis have been created in cigar‑shaped condensates, providing a practical route to the long‑junction geometry~\cite{rosenbusch2002dynamics}.
This extended core qualitatively reshapes the spatial structure of the weak link and, in turn, its transport properties, relative to the short-vortex case.
Considering three values of $a_{12}$, Fig.~\ref{fig:Long} examines how increasing the junction length modifies the CPR and the corresponding effective potential barrier.

Analogous to the short-junction case, at strong intercomponent repulsion [$a_{12}=220\,a_0$; Fig.~\ref{fig:Long}(a2)] the system lies firmly in the tunneling regime and displays an almost sinusoidal CPR (purple triangles).
Consistent with this, the effective barrier height for component 1 exceeds its chemical potential, $\mu_1$ [see inset of Fig.~\ref{fig:Long}(b), purple curve].
As the barrier height is reduced [see inset of Fig.~\ref{fig:Long}(b), other curves], salient differences from the short-junction case emerge.
For $a_{12}=150\,a_0$, the CPR supports stable solutions up to $\Delta\phi_1^{\mathrm{tot}}\approx3\pi/2$ [Fig.~\ref{fig:Long}(b); gold triangles], whereas for $a_{12}=130\,a_0$ it exhibits a zero-current stationary state at $\Delta\phi_1^{\mathrm{tot}}=2\pi$ [Fig.~\ref{fig:Long}(c2); green dots].
To interpret this intriguing behavior, we next turn to a circuit-model description.

\emph{Single- to double-junction crossover}---Applying the circuit model developed for the short vortex yields the dashed curves in Fig.~\ref{fig:Long}(a2,b,c2).
While this single-junction model reproduces the CPR in the tunneling regime [Fig.~\ref{fig:Long}(a2)], it fails dramatically at lower $a_{12}$ (lower barriers).
The breakdown is tied to the appearance of two density minima in component 1, separated by a local maximum within the elongated vortex core, due to long-range dipolar interactions [Fig.~\ref{fig:Long}(c1,c3)].
A further clue is that for $a_{12}=130\,a_0$ the current vanishes at $2\pi$ rather than at $\pi$.
Together, these observations indicate that, in the low-$a_{12}$ regime, the weak link effectively consists of two Josephson junctions in series, separated by a small central reservoir inside the vortex that contributes an additional kinetic inductance [schematic in Fig.~\ref{fig:Long}(c3)].
We evaluate this vortex inductance, $L_1^v$, from Eq.~\eqref{L_kin} by restricting the integration to the vortex-core region where $n_1(\rho=0,z)\ge n_1^{\mathrm{min}}+\big(n_1^{\mathrm{loc\,max}}-n_1^{\mathrm{min}}\big)/2$; see the two central vertical dashed lines in the schematic and the End Matter.
The resulting extended circuit model reads
\begin{equation}
\Delta\phi_i^{\mathrm{tot}}=\Delta\phi_i^{\mathrm{J1}}+\Delta\phi_i^{\mathrm{J2}}+\frac{(L_i+L_i^v)\,I_i}{\hbar},
\label{Eq:extendedmodel}
\end{equation}
where the vortex inductance substantially exceeds the outer inductance, for example, for $a_{12} = 130a_0$ we obtain $L_1^v=1.7\times 10^{-38}\,\mathrm{kg\,m^2}$ and $L_1=3.6\times 10^{-40}\,\mathrm{kg\,m^2}$.
By symmetry in the dc regime, the phase drops across the two junctions are identical, $\Delta\phi_i^{\mathrm{J1}}=\Delta\phi_i^{\mathrm{J2}}$.

Having established the double-junction picture, we benchmark it against full simulations.
The extended circuit model reproduces the CPRs in close agreement with the numerical results [solid lines in Fig.~\ref{fig:Long}(b,c2)].
We restrict its use to cases with two barrier maxima; in the tunneling regime at $a_{12}=220\,a_0$ the effective barrier has a single peak [inset of Fig.~\ref{fig:Long}(b)].
These comparisons show that long-range dipolar interactions reshape the vortex-core channel and enable a tunable crossover from a single junction to two coupled junctions as $a_{12}$ is reduced.

\emph{Conclusion}---We have shown that a vortex core acts as a self-induced weak link for superfluid transport, with an effective barrier set by the quantum pressure associated with the vortex-channel width.
This barrier---and hence the operating regime---is widely tunable via intercomponent interactions, relative component populations, and geometry, enabling a continuous crossover from hydrodynamic to tunneling transport.
Under long-range dipolar interactions, the core can be reshaped to realize a controlled transition from a single junction to two coupled junctions.
An extended Deaver-Pierce circuit model reproduces the numerically obtained CPRs and provides a framework for engineering multijunction atomtronic circuits in binary dipolar condensates.

Vortex-mediated weak links could be realized in uniform rings \cite{Polo2025} or in series with other junctions around a ring \cite{Pezz2024}, and as segments of cigar-shaped traps \cite{Li2022Long,Bland2022Alternating}, enabling studies of ac Josephson dynamics and complex quantum transport.
Similar junctions should be accessible in nondipolar mixtures \cite{myatt1997production,hall1998dynamics,maddaloni2000collective}, where the same control parameters provide practical tuning of barrier properties.

\emph{Acknowledgements}---We acknowledge thoughtful discussions with Klejdja Xhani, Giacomo Roati, and Ioan Pop. This research was funded in whole or in part by the Austrian Science Fund (FWF) [Grants DOIs 10.55776/COE1 and 10.55776/P36850]. For Open Access purposes, the author has applied a CC BY public copyright license to any author accepted manuscript version arising from this submission. The computational results presented here have been achieved (in part) using the LEO HPC infrastructure of the University of Innsbruck.

\appendix*

\onecolumngrid      

\section*{End Matter}

\twocolumngrid  

\setcounter{equation}{0}                         
\renewcommand{\theequation}{A\arabic{equation}}

\emph{Numerical details}---We consider a three-dimensional, binary dipolar condensate in an infinite-tube geometry aligned with the $z$ axis. The transverse harmonic trapping potential for each component is
\begin{equation}
V_i(\mathbf{x}) = \frac{m_i}{2}\bigl(\omega_x^2 x^2 + \omega_y^2 y^2\bigr),
\end{equation}
where $i=1,2$ indexes the components and $m_i$ denotes their masses.
At zero temperature, the dynamics are governed by two coupled dipolar Gross-Pitaevskii equations for the order parameters $\psi_i(\mathbf x,t)$,
\begin{align}
\label{SM:GPE}
i\hbar &\,\frac{\partial}{\partial t}\psi_i(\mathbf x,t)
= \Bigg[-\frac{\hbar^2}{2m_i}\nabla^2 + V_i(\mathbf x)
+ \sum_{j=1}^{2} g_{ij}\,n_j(\mathbf x,t) \nonumber\\
& + \sum_{j=1}^{2}\int d^3\mathbf x' \, U_{ij}(\mathbf x-\mathbf x')\,n_j(\mathbf x',t)\Bigg]\psi_i(\mathbf x,t),
\end{align}
where $n_i(\mathbf x,t) \equiv |\psi_i(\mathbf x,t)|^2$ is the atomic density of component $i$, and $g_{ij} = 2\pi\hbar^2 a_{ij}/m_{ij}$ are the contact interaction strengths, with $s$-wave scattering lengths $a_{ij}$ and reduced masses $m_{ij} = (m_i^{-1} + m_j^{-1})^{-1}$.
The long-range, anisotropic dipole-dipole interaction potentials are
\begin{equation}
\label{Eq:DDI}
U_{ij}(\mathbf r) = \alpha\,\frac{\mu_0\,\mu_i\,\mu_j}{4\pi}\,\frac{1 - 3\cos^2\theta}{r^3},
\end{equation}
where $\mu_i$ is the dipole moment of component $i$, $\mu_0$ is the vacuum permeability, $\mathbf r = \mathbf x - \mathbf x'$ is the relative position vector between two particles, $r = |\mathbf r|$, and $\theta$ is the angle between the $\mathbf r$ and the dipole polarization direction $\hat{\mathbf z}$.

While component 2 is nondipolar, we take the atoms of component 1 to rotate rapidly about the tube axis ($z$), resulting in an antidipolar situation [$\alpha = -1/2$ in Eq.~(\ref{Eq:DDI})].
Antidipoles energetically favor side-by-side dipolar arrangements, in contrast to the standard head-to-tail attraction.
Antidipoles have been realized using dipolar atoms \cite{Tang2018} and microwave-shielded polar molecules \cite{Will2025}.
We base our simulations on ${}^{162}\mathrm{Dy}$, with $m_1 = m_2 = 161.927\,\mathrm{u}$, but our findings are qualitatively applicable to a broad range of dipolar-nondipolar and dipolar-dipolar mixtures \cite{Trautmann2018,Ravensbergen2020,chalopin2020,Durastante2020,politi2022interspecies,Schafer2023,lecomte2025production,xie2025feshbach,kalia2025creation}.

\begin{figure}[t!]
	\centering
	\includegraphics[width=0.48\textwidth]{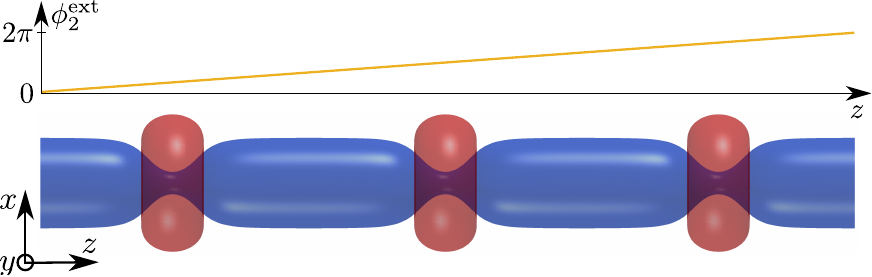}
	\caption{Protocol for imposing a linear phase gradient along the vortex-free component in a binary BEC. Imaginary time is then used to relax the phase profile to a DC stationary state.}
	\label{fig:EM_many}
\end{figure}

\emph{DC stationary states}---To obtain a stationary state with a centrally positioned vortex aligned along the trap axis ($z$), we exploit cylindrical symmetry and employ Fourier-Hankel transforms \cite{Ronen2006Bogolubov}.
This reduces the 3D problem to an effective 2D numerical treatment while enforcing the $2\pi$ phase circulation about the $z$ axis analytically.
The long-range dipole-dipole interactions are handled with a cylindrical cutoff \cite{Lu2010a}, truncating them in the radial direction to prevent unphysical interactions between radial Fourier copies.
For antidipolar-nondipolar mixtures in the immiscible phase, the ground state consists of alternating domains, cf.~Fig.~\ref{fig:EM_many} (see also \cite{Kirkby2023Spin}).
We initially consider a single unit cell containing one full domain of each component.
Starting from a state with a vortex imprinted in one component, we evolve it with Eq.~\eqref{SM:GPE} in imaginary time using a split-step Fourier method.
Along $z$, Fourier copies are allowed to interact by choosing the cutoff length much larger than the characteristic size of any density modulation, thereby realizing the infinite-tube geometry from the perspective of interactions.
For each parameter set, we determine the unit cell length $l$ that minimizes the energy per particle, enabling us to consider a single, energetically optimal unit cell (referred to in the main text as the system length $l$) \cite{blakie2020supersolidity,Smith2023}.

 \begin{figure}[b!]
	\centering
	\includegraphics[width=0.48\textwidth]{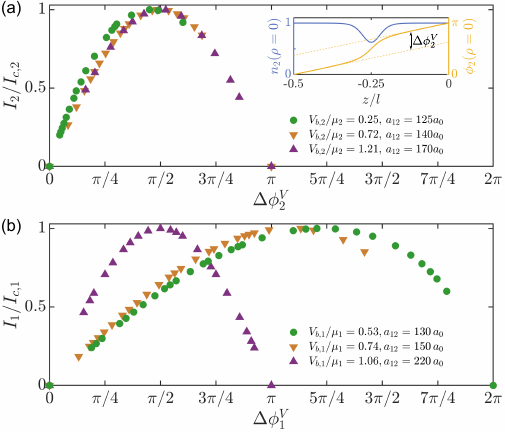}
	\caption{CPRs for (a) the short vortex and (b) the long vortex, plotted as a function of the phase difference $\Delta\phi_i^V$ restricted to the vortex core region. The inset in (a) illustrates the localized phase drop across the junction.}
	\label{fig:EM_CPR_j}
\end{figure}

To obtain stationary states carrying core currents, we start from a converged ground state and impose a linear phase gradient along the axial ($z$) direction in the component without a vortex.
This is achieved by multiplying the corresponding wave function by a phase factor $\exp[i\,\phi_i^{\rm ext}(z)]$ with $\partial_z \phi_i^{\rm ext}=\mathrm{const}$.
For a single unit cell, the imposed phase difference must be an integer multiple of $2\pi$ due to periodic boundary conditions.
To realize arbitrary phase differences, we extend the system to multiple unit cells: the phase difference per unit cell is then $2\pi q/n_{\rm uc}$, where $q\in\mathbb{Z}$ and $n_{\rm uc}\in\mathbb{Z}$ is the number of unit cells.
For example, to impose a phase difference of $\Delta\phi_i^{\rm tot}=2\pi/3$ per unit cell, we simulate three unit cells with $\int_{-l/2}^{l/2}\partial_z \phi_i^{\rm ext}(z)\,dz = 2\pi$, as schematically depicted in Fig.~\ref{fig:EM_many}.
We then evolve the coupled Gross-Pitaevskii equations~\eqref{SM:GPE} in imaginary time, allowing the system to relax into a stationary state that carries a quantized current.
This procedure enables the computation of current--phase relations (CPRs) for both short and long vortex configurations.

\emph{Phase drop through vortex channel}---In addition to the CPRs shown in the main text (Figs.~\ref{fig:Short} and \ref{fig:Long}), we present complementary CPRs from the same data that consider only the phase drop across the vortex channel [see inset of Fig.~\ref{fig:EM_CPR_j}(a)].
The CPRs for the short vortex in Fig.~\ref{fig:EM_CPR_j}(a) exhibit the classic sinusoidal behavior, indicating that the inductance is spatially separated from the ideal junction region within the vortex core.
In contrast, the long-vortex junction accommodates phase differences exceeding $\pi$ [Fig.~\ref{fig:EM_CPR_j}(b)], supporting the interpretation that the long vortex behaves as a double-junction.

\begin{figure}[t!]
	\centering
	\includegraphics[width=0.48\textwidth]{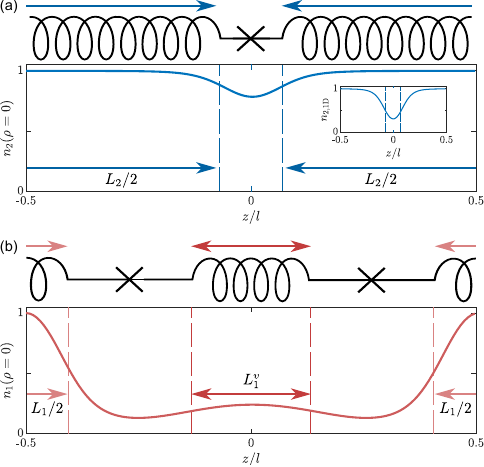}
	\caption{
		Identification of junction regions.
		(a) Three-dimensional density along the vortex core for a short-vortex configuration ($a_{12}=125\,a_0$).
		Dashed lines delineate the junction region and the segment used to evaluate the kinetic inductance. The inset shows the one-dimentional density along the vortex core.
		(b) For the long vortex ($a_{12}=130\,a_0$), the local density maximum within the core acts as an additional small reservoir, contributing an extra kinetic inductance $L_1^{v}$. Densities are normalized to their respective peak values.
	}
	\label{fig:EM_induc}
\end{figure}

\emph{Numerical calculation of kinetic inductance}---To evaluate the kinetic inductance outside the junction region, we first determine the junction boundaries from the three-dimensional density of the flowing component along the vortex core, $n_i(\rho=0,z)$.
We identify points belonging to the junction via the criterion
$n_i(\rho=0,z) < n_i^{\min} + f\big(n_i^{\max} - n_i^{\min}\big)$,
where $n_i^{\min}$ and $n_i^{\max}$ are the global minimum and maximum of $n_i(\rho=0,z)$, and $f=0.5$ is a chosen threshold coefficient.
An example illustrating the segregated junction region for a short-vortex case is shown in Fig.~\ref{fig:EM_induc}(a).

For the long vortex, the long-range dipole-dipole interactions within component 1 modify the core structure and create a local density maximum inside the vortex core [Fig.~\ref{fig:EM_induc}(b)].
Consequently, the long-vortex core effectively behaves as two junctions separated by an additional kinetic inductance $L_1^{v}$ associated with this local maximum [see Fig.~\ref{fig:EM_induc}(b)].
To evaluate this additional inductance, we apply the same criterion as above, but replace the global maximum by the value of the local maximum.

\end{document}